\documentclass{aa}
\usepackage{graphicx}

\begin{document}

\sloppypar

   \title{Broadband X-ray spectrum of intermediate polar V1223 Sgr}

   \author{M. Revnivtsev \inst{1,2}, A. Lutovinov \inst{1}, V. Suleimanov \inst{3}, R. Sunyaev\inst{1,2}, V. Zheleznyakov\inst{4}}

   \offprints{revnivtsev@hea.iki.rssi.ru}

   \institute{Space Research Institute, Russian Academy of Sciences,
              Profsoyuznaya 84/32, 117810 Moscow, Russia
        \and   
                Max-Planck-Institute f\"ur Astrophysik,
              Karl-Schwarzschild-Str. 1, D-85740 Garching bei M\"unchen,
              Germany
        \and   
                Kazan State University, Kremlevskaya 18, 420008, Kazan, Russia
	\and
	      Institute of Applied Physics, Ulyanova 46, 603950 Nizhny 
              Novgorod,Russia
            }
  \date{}

        \authorrunning{Revnivtsev et al.}
        \titlerunning{}
 
   \abstract{We present the broadband phase averaged 
spectrum of one of the brightest
intermediate polars V1223 Sgr, obtained with INTEGRAL and RXTE 
observatories (3-100 keV). Good statistical quality of the spectrum in
a hard X-ray energy 
band (INTEGRAL/IBIS and RXTE/HEXTE) allowed us to disentangle
contributions of a direct optically thin plasma emission and a 
reflected component to the spectrum of V1223 Sgr. The obtained measurement 
of the post-shock temperature of the accreting matter give us the
information about the mass of the white dwarf and the inclination 
of the system. 
   }

   \maketitle
%
%________________________________________________________________

\section{Introduction}

V1223 Sgr is a very well known binary star that belongs to 
the class of intermediate polars (IPs) -- binary systems with a white
dwarf that accrets matter from a Roche lobe filling secondary
star (see e.g. comprehensive model and discussion in \cite{beuermann04}). 
In the IP systems magnetic field of a white dwarf is strong enough to control
the accreting matter at relatively large distances from it and to form
a magnetosphere and magnetic funnel (see e.g. \cite{patterson94} for a review).
The matter falls down to the surface of the white dwarf with
almost free fall velocity and is heated in a strong shock to temperatures
of the order of $\sim 10$ keV. Falling matter preferentially comes to the
magnetic poles and emits X-rays via an optically thin thermal plasma 
emission (see e.g. \cite{lamb79}).
In spite of this relatively simple scheme the observed emission of IP
is much more complicated. First of all its spectrum can contain a set of 
components with different temperatures (mostly visible at low energies),
it often demonstrates an intrinsic photoabsorption, partial
covering, different ionization states of an emitting plasma and a 
reflection from an optically thick relatively medium (see e.g. 
\cite{warner95} for a review). 
Often the spectrum depends on orbital phase and phase of rotation of the white 
dwarf (e.g. \cite{norton89}).
Different objects of this class were extensively studied by several X-ray
 observatories over
last decades. However, until now there was no study of hard X-ray 
emission of IP ($E>20$ keV) except for the case of V709 Cas (de Martino et
al. 2001). 

Energy band $>$ 20 keV is important because it allows
to confidently measure the maximal temperature of the shock in the accretion
column of the IP. It is especially important if the X-ray
spectrum of IP contains strong reflected component, that distorts the observed 
continuum at energies $\sim$10-20 keV. In the most cases the temperature
of the post-shock region of the accretion column of 
IP is of the order of $\sim 10-40$ keV
that makes it quite difficult to measure with telescopes working
at 0.5-10 keV energy band, like EINSTEIN, ASCA, CHANDRA, XMM.  
Observations of IP performed with GINGA/LAC (e.g. Ishida 1991, 
Ishida et al. 1994) 
practically
do not provide statistically significant measurements at  $E\sim$20 keV, that 
makes it hard to confidently measure the temperatures higher than 
$\sim 10-20$ keV. As the energetics of the X-ray spectrum of IP is dominated 
by bremsstrahlung emission with maximal temperatures the accuracy of its
measurement is of high importance (see. e.g.
discussion in \cite{beuermann04}).

In this work we present results of measurements of the broadband spectrum
of the intermediate polar V1223 Sgr with the RXTE and INTEGRAL observatories
with an accent on the averaged hard X-ray part of the spectrum.

\section{Data analysis}

The international gamma-ray observatory INTEGRAL was launched by the Russian 
launcher PROTON from the Baikonur cosmodrome in the high-apogee orbit 
on October 17, 2002 (Eismont et al. 2003). The payload includes four 
principal instruments which allow one to carry out simultaneous observations of
sources in  
the X-ray, gamma-ray and optical energy range (Winkler et al. 2003). 

In the August-September of 2003 ultra deep ($\sim$2 million sec)
observations of the Galactic center region were performed. 
The survey of this region with catalog of detected sources
was published by Revnivtsev et al. (2004). The intermediate polar
V1223 Sgr was detected during these observations 
with a high statistical significance in the 18-60 keV energy band.

Due to relatively large distance from the Galactic Center 
($\sim 16^\circ$), V1223 Sgr was outside the field of view of the
X-ray monitor JEM-X of the INTEGRAL observatory. On the other hand the source
is weak and relatively soft, therefore the sensitivity of the spectrometer
SPI (\cite{spi}) is not enough to measure the spectrum of the source. 
Therefore in the paper we will concentrate
on the data obtained  by IBIS
telescope, its upper detector ISGRI (\cite{lebrun03}). 
The source was quite far from the center of the field of view of
the IBIS telescope, resulting in a strong (factor of $\sim$9.5) 
reduction of its effective area useful for the measurement of the 
spectrum of V1223 Sgr.

\begin{figure}
\includegraphics[width=\columnwidth]{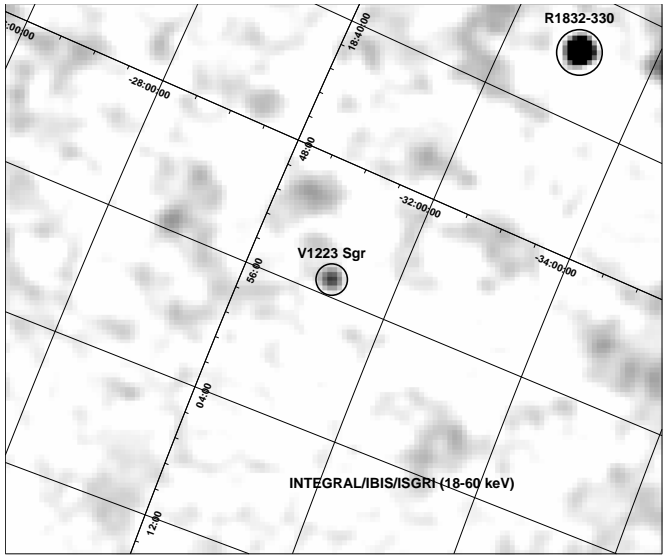}
\caption{X-ray map of the region around V1223 Sgr obtained with 
INTEGRAL/IBIS/ISGRI in the energy band 18-60 keV}
\end{figure}

Data of the IBIS telescope were processed using methods described in
Revnivtsev et  
al. (2004). The image of the region around V1223 Sgr is presented in Fig.1.
For the spectral analysis we used a simple ratio of the flux
measured from the source to the fluxes of the Crab nebula in the same
energy bands and assumed that the Crab nebula spectrum has the form 
$d N(E)=10 E^{-2.1} dE$, where $N(E)$ is the number of photons at energy $E$.
 As the spectrum of V1223 Sgr does not demonstrate
very sharp structures such method of the spectral reconstruction is
acceptable. Tests of this method on observations of the Crab nebula
showed that the resulted spectra can still possess approximately 
5\% systematic uncertainties in every energy channel.

In order to compare the obtained results with measurements of other instruments
and to accomplish the obtained hard X-ray spectrum of the source 
with standard X-ray band measurements ($\sim$1--20 keV) we used 
data of the RXTE observatory (\cite{rxte}, \cite{rothschild98}). 
The PCA and HEXTE spectrometers of this
observatory cover the energy range 3--250 keV. Effective areas
of these spectrometers are quite large $\sim 6400$ cm$^2$ for the PCA
at energies $\sim$6-7 keV and $\sim$650 cm$^2$ for each (of two) 
clusters of HEXTE detectors at energies $\sim$50 keV. We quote here
effective area of only one HEXTE cluster of detectors because only one
HEXTE cluster can observe the source at the same time, the
other one observes the background at this time.

Intermediate polar V1223 Sgr was observed by the RXTE observatory 14 times
during period 1996-2000. Total usable exposure time of these observation
is approximately 114 ksec.

The data of the RXTE observatory was reduced with the help of standard 
LHEASOFT/FTOOLS 5.3 package. As the spectrum of IPs does not vary much at 
energies higher than $>$3 keV (see e.g. \cite{beardmore00} for the 
case of V1223 Sgr) for the following analysis 
we averaged all available data and analyzed only averaged spectrum of V1223 
Sgr. For the spectral approximation we used XSPEC package.
Normalizations of spectra obtained by different instruments were set free
to vary during the fitting procedure. Systematic errors 1\% and 5\% were
added to every channel of the RXTE/PCA and INTEGRAL/IBIS spectra,
respectively. The resulted spectra of all 
instruments were rescaled to that of the RXTE/PCA.

\section{Spectral analysis}

First of all we approximated the hard X-ray part of the spectrum of V1223 Sgr
obtained with INTEGRAL/IBIS and RXTE/HEXTE  (energy band 20-100 keV) 
with a simplest thermal model -- an
optically thin thermal bremsstrahlung. The quality of the fit is very good
(reduced $\chi^2\sim 0.9$) and the parameter of the temperature
is equal to $kT=29\pm2$ keV. Best fit parameters for separate data of 
RXTE/HEXTE and INTEGRAL/IBIS are $25.6\pm 2.5$ keV and $29\pm 5$ keV, 
respectively. As the obtained results agree well with each 
other in our subsequent analysis we will combine data of these two 
instruments. Obtained value of the temperature 
parameter is somewhat smaller than that obtained by Beardmore et al. 
(2000) from GINGA data ($kT=43\pm13$ keV).
However if we apply the same model (with the neutral absorption 
and broad gaussian line in the energy range $\sim 6-7$ keV) to the data of the RXTE/PCA detector
(energy band 3--20 keV) not taking into account harda X-ray data
 the resulted temperature would be in a very good 
agreement with that of Beardmore et al. (2000) -- $kT=44.5\pm1.5$ keV. 
Moreover, if we will try to approximate only spectral points in the energy band 10--20 keV we will obtain even higher temperature -- $kT=65 \pm 5$.
This discrepancy can be understood if we will take into account that 
the spectrum of V1223 Sgr contains relatively strong reflected (e.g. \cite{basko74}) component
and therefore the model for the spectral approximation should be more complex.

In the work of Beardmore et al. (2000) only GINGA/LAC (energy range 
2--37 keV) data were used to describe the source spectrum at energies 
$>$10 keV. However, the spectrum of V1223 Sgr obtained by GINGA/LAC
practically does not contain statistically significant points at energies 
higher than $\sim 20$ keV . In spite of the fact that 
Beardmore et al. (2000) included reflected component into their spectral 
approximation it is very hard for their dataset to determine at the same time
the temperature of the thermal plasma and the amplitude of the reflection.
In our case we have data of INTEGRAL/IBIS and the RXTE/HEXTE (20-100 keV) 
and can determine this parameters more confidently while fitting the broadband
source spectrum. 

Approximation of the broadband spectrum (BeppoSAX data) 
of the another intermediate polar V709 Cas made by de Martino et al. (2001) confirms
these arguments. Authors showed that the broadband spectrum of IP
should be described by a thermal model with the reflection. Neglecting the 
X-ray reflection in a spectral modeling result in the overestimation of the temperature
of a post-shock plasma in the accretion column of IP.

\begin{figure}
\includegraphics[width=\columnwidth,bb=20 180 566 710,clip]{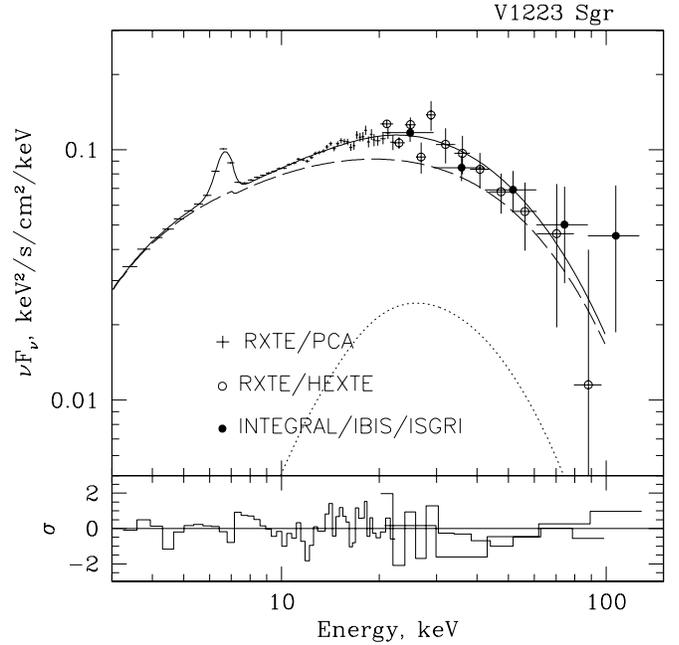}
\caption{Averaged spectrum of V1223 Sgr obtained with INTEGRAL and RXTE 
observatories. Solid curve represents the best fit model. Dashed line
shows the emission of the thermal plasma, dotted line -- 
the contribution of the reflected component. On the lower panel the
difference between data and the model in the units of standard deviations
($\sigma$) is shown. }
\end{figure}

\begin{figure}
\includegraphics[height=\columnwidth,bb=45 40 570 700,clip,angle=-90]{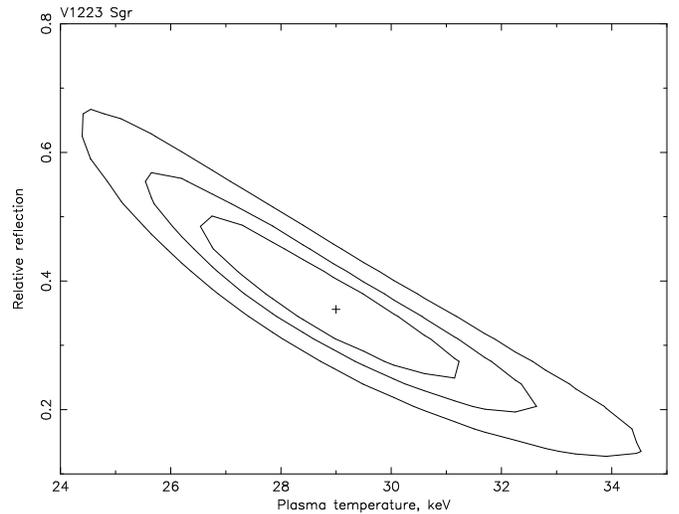}
\caption{Confidence contours (1,2 and 3 $\sigma$) of the plasma temperature ($kT$) and relative reflection amplitude ($R=\Omega/2\pi$)}
\end{figure}

\begin{table}
\caption{Best fit parameters for averaged broadband (3-100 keV) 
spectrum of V1223 Sgr}
\tabcolsep=7mm
\begin{tabular}{l|l}
Parameter&\\
\hline
$N_{\rm H}$, $10^{22}$ cm$^{-2}$&$3.3\pm0.2$\\
$kT$, keV&$29\pm2$\\
$R=\Omega/2\pi$&$0.35\pm 0.15$\\
$E_{\rm gauss}$&$6.63\pm 0.03$\\
$\sigma_{\rm gauss}$&$0.37\pm0.02$\\
$EW$, eV&$387\pm18$\\
Flux$_{\rm obs}$, $10^{-10}$ erg/s/cm$^2$&$3.76\pm0.04 ^a$\\
Flux$_{\rm corr}$, $10^{-10}$ erg/s/cm$^2$&$4.57\pm0.05 ^b$,\\
$\chi^2$/d.o.f.&76.5/89\\
\hline
\end{tabular}
\begin{list}{}
\item -- $R=\Omega/2\pi$ -- is the parameter of the reflection, solid angle
  subtended by a reflector 
\item $^a$ -- observed flux in the energy band 3-100 keV, the 
flux was calculated assuming that the spectrum of the Crab nebula
has the following normalization $dN(E)=10.0 E^{-2.1} dE$. If the assumption 
of the Crab nebula spectrum is different the flux value should be rescaled
\item $^b$ -- absorption corrected flux in the energy band 0.5-100 keV
\end{list}
\end{table}

For the broadband spectral approximation we used a simple analytic model
consisting of the bremsstrahlung emission with its reflection 
from an optically thick cold medium plus the neutral absorption 
and broad gaussian line at $\sim$6.5 keV. The latter component  mimics 
a set of lines of optically thin plasma and 
fluorescent Fe line, originating as a result of the reflection.
Such spectral model should approximately represent
emission of the hot plasma in the post-shock region of V1223 Sgr. 
The best fit temperature of the bremsstrahlung emission in this case will
denote the maximal temperature in the post-shock region of the 
accretion column of IP. Obtained best fit parameters of the approximation
of the combined INTEGRAL and RXTE spectrum of V1223 Sgr are presented in
Table 1. 
The averaged spectrum of V1223 Sgr is shown in Fig. 2. 
As the value of the plasma temperature strongly correlates with the 
amplitude of the reflection we also present confidence contours on these
values in Fig. 3.

\section{Discussion}

In the previous section we analyzed the broadband phase averaged spectrum 
of one of the brightest intermediate polar V1223 Sgr. We have shown
that even apart from detection of fluorescent Fe line at energy $\sim 
6.4$ keV (see \cite{beardmore00}) there are strong indications on 
the presence of the reflection component in the spectrum of V1223 Sgr. 
Because of this component
any measurements of the post-shock temperature of the accretion column 
near the surface of IP that are based on instruments with an effective 
energy band lower than $\sim$20-30 keV 
are biased towards obtaining higher temperatures.
Our spectral modeling for the first time includes hard X-ray energy range 
(20-100 keV) and
provides more confident estimation of the post-shock temperature. 
Using this value we derive some important parameters
of V1223 Sgr.

Current models of a shock in the accretion column of IP assume
that the accreted matter is heated up to its virial temperature
(see e.g. \cite{aizu73}, \cite{fabian76}, \cite{lamb79}).
The heated plasma produces mainly bremsstrahlung with possible
contribution from a cyclotron emission at low accretion rates
(see e.g. \cite{fabian76}, \cite{warner95}). 

Our new broadband measurements of the X-ray spectrum of V1223 Sgr 
provide us a set of observables  with
higher confidence than that  
of previous. They are: 1) the post-shock temperature $T_{\rm s}=29\pm2$ keV, 
2) the broadband X-ray flux (absorption corrected) 
$F_{\rm 0.5-100 keV}=4.6\times10^{-10}$ erg
 s$^{-1}$ cm$^{-2}$, that in turn gives us the emission measure of one 
emitting accretion column
$E_{\rm m, 1} =\int N_e^2 dV$ = $1.07\times 10^{13}~4\pi D^2$ cm$^{-3}$, where $N_e$ -- is the number density of electrons in a hot emitting plasma.

   The distance to V1223 Sgr is now known from HST measurement of the parallax
of the system, $D=527 ^{+54}_{-43}$ pc (Beuermann et al. 2004).
 Therefore we can
evaluate the isotropic luminosity of one accretion column $L_{\rm x,1} 
= 4\pi D^2 F_{\rm 0.5-100 keV}$ = 
$(1.5 \pm 0.3)  \times 10^{34}$ erg s$^{-1}$, 
and the volume emission measure E$_{\rm m,1} \approx  3.4 \times 10^{56}$
cm$^{-3}$. 

The shock temperature can give the direct evaluation of a white dwarf
mass (see e.g. \cite{fabian76}, \cite{warner95})

\[
   T_s = 31.9 \frac{M_{\rm WD}}{M_{\odot}} \frac{10^9 \rm cm}{R_{\rm WD}} \rm keV.
\]

Using an approximation of the white dwarf mass-radius relation
from Nauenberg (1972) 

\[
R_9=0.78\left[\left({1.44 M_\odot\over{M}}\right)^{2/3}-\left({M\over{1.44 M_\odot}}\right)^{2/3}\right]^{1/2}
\]

where $R_9$ -- is the radius of the white dwarf in the units of $10^9$ cm, 
we obtain for the mass and radius of the white dwarf in the V1223 Sgr:

\[
  M_{\rm WD}= 0.71 \pm 0.03 M_{\odot}
\]
\[
R_{\rm WD}=(7.8\pm0.2)\times 10^8 \rm cm
\]

The obtained estimation of the white dwarf mass is slightly smaller than
that obtained by Beardmore et al. (2000)
because of lower post-shock
temperature measured by us 
and is better consistent with the 
evaluation of white dwarf mass from optical observations: 
$M_{\rm WD}=0.4-0.6 M_\odot$ (\cite{penning85}).

The mass function of the secondary star in the system V1223 Sgr is 
$f(M_2)=0.0026 \pm 0.0015$
 $M_{\odot}$ (Watts et al. 1985), the mass of the secondary star is
 $M_{2}$ = 0.4 $M_{\odot}$ (Penning 1985). Using these values
we can obtain the inclination angle of the system $i = 22^\circ \pm
5^\circ$. If we will assume that the mass of the
secondary star is lower, e.g. $M_{2}=0.25 M_{\odot}$ (\cite{beuermann04})
the estimation of the inclination angle will increase to $i=32^\circ
\pm 7^\circ$.

As the X-ray emission is the major channel of energy losses in the post-shock
region we can derive the mass accretion rate $\dot M$

\[
    \dot M \approx {3\over{8}} {\mu m_{\rm H} 2 L_{\rm x,1} \over{k T_{\rm s}}}
\]
where $\mu$=0.62 is the mean molecular weight of the accreting matter.
Here factor 2 stands in order to account for two emitting magnetic
poles of IP. Substituting measured values we obtain
$\dot M \approx 2.54 \times 10^{17}$ g s$^{-1}$.

Using $M_{\rm WD} = 0.71 M_{\odot}$, $R_{\rm WD} = 7.8 \times 10^8$ cm 
and $E_{\rm m,1} = 4 \pi R^2_{\rm WD} f/2 ~ h_s N_e^2 \sim 
3.4 \times 10^{56}$ cm$^{-3}$, where $f$ -- is the fraction of the white 
dwarf surface, occupied by two accretion columns,
 we can evaluate the height of 
standing shock $h_{\rm s}$ and the number density of electrons
in the post-shock region $N_e$ following Fabian et al. (1976) and
 Warner (1995).

\[
     N_e = 3.1 \times 10^{13} \dot M_{17} \left(\frac{M_{\rm WD}}{M_{\odot}}\right)^{1/2} R_{9}^{-3/2} f^{-1}
\]
\[
     h_s = 3.3 \times 10^{4} f \,\, \dot M_{17} \left({M_{\rm WD}\over{M_\odot}}\right)^{3/2} R_9^{1/2} \rm cm
\]
where $\dot M_{17}$ -- is the mass accretion rate in units of $10^{17}$ g 
s$^{-1}$. We can derive
$N_e=9.6\times10^{16} f^{-1}_{-3}$ cm$^{-3}$ and 
$h_{\rm s}=1.1 \times 10^{7} f_{-3}$ cm, where $f_{-3}=10^{-3} f$ -- is the 
fraction of the surface of the white dwarf, occupied by two accretion
columns  
in the units of $10^{-3}$. Note that the derived parameters are similar to 
those obtained by de Martino et al. (2001) for the
intermediate polar V709 Cas.

\begin{acknowledgements}
Authors thank Eugene Churazov for developing of algorithms of 
analysis of data of IBIS telescope and for providing the software.
Research has made use of data obtained 
through the INTEGRAL Science Data Center (ISDC), Versoix, Switzerland, 
Russian INTEGRAL Science Data Center (RSDC), Moscow, Russia,  
and  High Energy Astrophysics Science Archive Research Center 
Online Service, provided by the NASA/Goddard Space Flight Center.
Work was partially supported by grants of
Minpromnauka NSH-2083.2003.2 and 40.022.1.1.1103 and program of Russian
Academy of Sciences ``Non-stationary phenomena in astronomy''.
MR and AL thank International Space Science Institute (ISSI), Bern for
hospitality 
and partial support of the work.
VS acknowledge the support of RFFI grant 02-02-17174 and  
program of the president of Russian Federation 
of support of scientific schools NSH-1789.2003.2

\end{acknowledgements}

\end{document}